\documentclass[conference]{IEEEtran}
\IEEEoverridecommandlockouts

\usepackage{cite}
\usepackage{amsmath,amssymb,amsfonts}
\usepackage{algorithmic}
\usepackage{graphicx}
\usepackage{textcomp}
\usepackage{xcolor}
\usepackage{subfigure}
\usepackage{geometry}

\begin{document}

\title{A Low-Complexity Range Estimation with Adjusted Affine Frequency Division Multiplexing Waveform}

\author{\IEEEauthorblockN{Jiajun Zhu\textsuperscript{1}, Yanqun Tang\textsuperscript{1*}, Xizhang Wei\textsuperscript{1}, Haoran Yin\textsuperscript{1}, Jinming Du\textsuperscript{1}, Zhengpeng Wang\textsuperscript{1}, Yuqinng Liu\textsuperscript{2}}
\IEEEauthorblockA{\textit{\textsuperscript{1}School of Electronics and Communication Engineering, Sun Yat-sen University, Shenzhen, China} \\
\textit{\textsuperscript{2}School of Automation and Electronic Information, Xiangtan University, Xiangtan, China}\\
Email: zhujj59@mail2.sysu.edu.cn, \textsuperscript{*}tangyq8@mail.sysu.edu.cn}
}

\maketitle

\begin{abstract}
Affine frequency division multiplexing (AFDM) is a recently proposed communication waveform for time-varying channel scenarios. As a chirp-based multicarrier modulation technique it can not only satisfy the needs of multiple scenarios in future mobile communication networks but also achieve good performance in radar sensing by adjusting the built-in parameters, making it a promising air interface waveform in integrated sensing and communication (ISAC) applications.
In this paper, we investigate an AFDM-based radar system and analyze the radar ambiguity function of AFDM with different built-in parameters, based on which we find an AFDM waveform with the specific parameter $c_2$ owns the near-optimal time-domain ambiguity function.
Then a low-complexity algorithm based on matched filtering for high-resolution target range estimation is proposed for this specific AFDM waveform.
Through simulation and analysis, the specific AFDM waveform has near-optimal range estimation performance with the proposed low-complexity algorithm while having the same bit error rate (BER) performance as orthogonal time frequency space (OTFS) using simple linear minimum mean square error (LMMSE) equalizer.
\end{abstract}


\section{Introduction}
Integrated sensing and communication (ISAC) is a novel technology that combines radar sensing and data communication with the goal of realizing the coexistence of frequency bands, the sharing of hardware and algorithms, and the synergistic gain between them, and it has thus been established as one of the visions for future sixth-generation wireless communication networks \cite{6G}.
One of the key technologies in ISAC is the communication-centered ISAC waveform design, which integrates the function of radar sensing into the communication waveform to assist communication by improving the effectiveness and reliability performance \cite{ISAC1,ISAC2}. Therefore, before realizing the synergy between them, the radar sensing performance of the waveform must be studied, and a portion of the radar detection function must be realized.

Orthogonal time frequency space (OTFS) is a communication waveform proposed by \emph{Hadani} in 2017 with high robustness to Doppler frequency deviation \cite{OTFS}. It can capture the full diversity gain of data communication based on signal processing in the delay Doppler domain, and, due to the linkage of delay and Doppler to the range and velocity of the radar target, OTFS naturally link communications theory and radar theory and has become one of the most promising waveforms that are focused on in ISAC applications \cite{DDRADAR,OTFSISAC}.
The ambiguity function is a key metric of radar sensing performance, and it can directly measure the range and velocity estimation resolution of pulse regime radars \cite{ABF}. However, due to the poor shape of the time-domain ambiguity function of OTFS, it is impossible to use the classical pulse compression (i.e., an algorithm based on matched filtering) algorithm in the time domain directly \cite{OTFSABF}.

Recently, affine frequency division multiplexing (AFDM) has been proposed as a multicarrier modulation technique based on linear frequency modulation (LFM, i.e., chirp) signals. Similar to OTFS, it obtains the full diversity gain by extending the modulation symbols throughout the entire time-frequency plane and thus is also a suitable communication waveform for time-varying channel scenarios \cite{AFDM2,AFDM}. Furthermore, since chirp signals are classical detection waveforms with high performance in radar applications, AFDM naturally connects the fields of communication and radar, making it one of the most promising waveforms for ISAC applications \cite{AFDMISAC1}.
In \cite{AFDMISAC2}, an AFDM-based ISAC system was proposed with a time-domain parameter estimation scheme utilizing the fast cyclic correlation radar algorithm. However, the multiple chirp-periodic prefixes (CPP) of multiple AFDM symbols greatly increase the transmission time of the waveform, which can lead to slower tracking rates as well as shorter detection ranges.

In this paper, we consider a single AFDM symbol for both bistatic and monostatic radar sensing applications.
The effect of AFDM built-in parameters on its ambiguity function is first investigated, and we find the specific AFDM waveform with parameter $c_2=0$ owns the near-optimal time-domain ambiguity function, i.e., narrower main lobe and lower side lobe.
Based on this waveform, a simple matched filter-based algorithm in the time domain can achieve high resolution range estimation.
Furthermore, the reasonable parameter selection degrades the CPP into a simple cyclic prefix (CP) similar to that used in orthogonal frequency division multiplexing (OFDM). Based on the cyclic nature of the CP, we propose a frequency-domain parameter estimation method utilizing the fast Fourier transform (FFT) to further reduce the computational complexity of the matched filter-based algorithm.

As a result, the single-symbol AFDM waveform with selected parameters not only has the near-optimal time-dimensional ambiguity function similar to OFDM, but also inherit the ability of time-frequency diversity similar to OTFS to adapt to time-varying environments. More importantly, the short time waveform and simple ranging algorithm based on matched filtering proposed in this paper can greatly improve signal processing computational efficiency to meet the real-time range estimation requirements.
Through simulation analysis, the specific AFDM not only possesses more accurate ranging resolution and noise immunity compared to OTFS but also maintains the same BER performance in a time-varying channel (moving radar targets) environment.


\begin{figure}[t]
	\centering		\includegraphics[width=1\linewidth,scale=1.00]{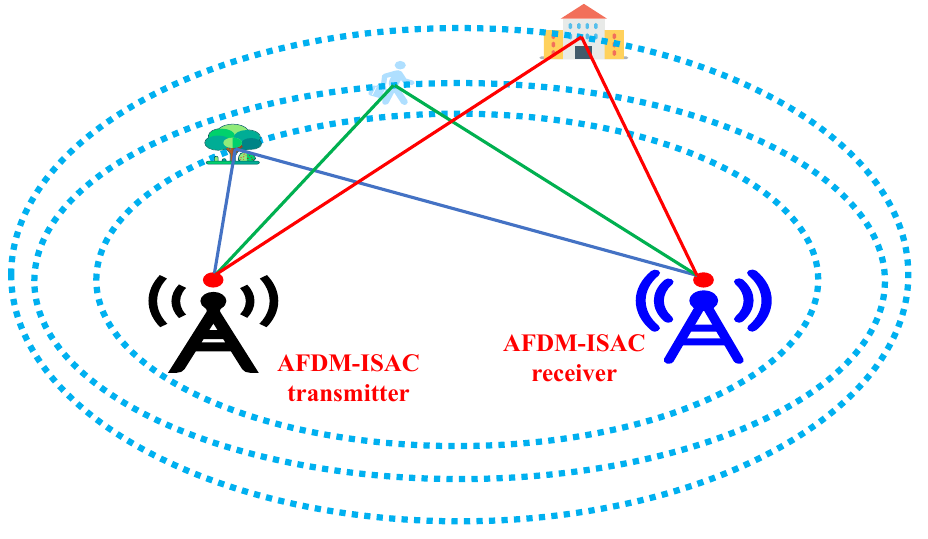}	
	\caption{AFDM-based bistatic radar for range estimation.}	
	\label{AFDM-ISAC-range}	
\end{figure}

\section{AFDM-BASED RADAR SYSTEM MODEL AND PERFORMANCE METRIC}
\subsection{AFDM-Based Radar Sensing Signal Model}

Considering an environment consisting of multiple moving scatterers in a bistatic radar system, we have its continuous time expression
\begin{equation}
g(t,\tau)=\sum_{p=1}^P h_p e^{j2\pi \nu_p(t-\tau)}\delta (t-\tau_p), \label{TDchannel}
\end{equation}
where $\delta(\cdot)$ is the Dirac delta function, $P$ is the number of scatters (targets), $h_p$, $\tau_p$ and $\nu_p$ denote the complex gain, delay, and Doppler frequency of the $p$-th target, respectively. For bistatic radar sensing, the delay between the transceiver and the $p$-th target is given by
\begin{equation}
\tau _p=\frac{R_{p}^{\mathrm{t_x}}+R_{p}^{\mathrm{r_x}}}{V_\mathrm{c}},\label{range}
\end{equation}
where $V_\mathrm{c}$ is the speed of light, $R_{p}^{\mathrm{t_x}}$ and $R_{p}^{\mathrm{r_x}}$ are the $p$-th target's range relative to the transmitter and receiver (note that since the distance between the transmitter and receiver is easy to know, we only consider the relative delay of the other paths to the direct path of the transceiver), respectively. Fig. \ref{AFDM-ISAC-range} is a schematic diagram of bistatic radar ranging, and it should be noted that \eqref {range} is also applicable to monostatic radars when $R_{p}^{\mathrm{t_x}}=R_{p}^{\mathrm{r_x}}$.

\begin{figure}[t]	
	\centering		\includegraphics[width=\linewidth,scale=1.00]{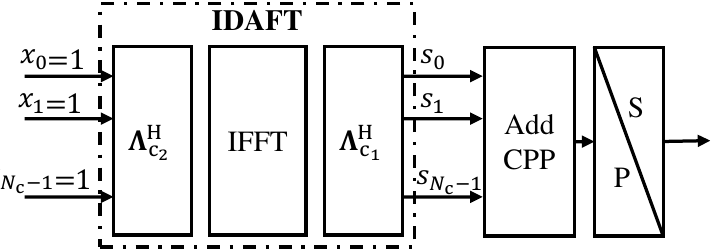}	
	\caption{Block diagram of AFDM waveform without modulating information.}	
	\label{AFDM-modulate}	
\end{figure}

Considering an AFDM symbol with a duration of $T$ and bandwidth of $B_\mathrm{c}=N_\mathrm{c}\Delta f$, where $\Delta f=1/T$ denotes the chirp subcarrier spacing, we can obtain the delay and Doppler indices by sampling according to the delay resolution of $1/B_\mathrm{c}$ and Doppler resolution of $1/T$ as
\begin{equation}
{l_p}={N_\mathrm{c}\Delta f}\tau_p\quad, \quad {k_p+\kappa_p}={T}\nu_p,\label{rangesample}
\end{equation}
where $l_p$ and $k_p$ are the delay and Doppler indices of the $p$-th target, the term $\kappa_p\in(-0.5,0.5)$ is the fractional shifts from the nearest Doppler index. Specifically, we have $0 \le l_p \le l_{p,\mathrm{max}}$, where $l_{p,\mathrm{max}}$ is the maximum delay index satisfying $l_{p,\mathrm{max}} = N_\mathrm{c} \Delta f \tau_{p,\mathrm{max}}<N_\mathrm{c}$, and $\tau_{p,\mathrm{max}}=V_\mathrm{c}T/2$ is the maximum unambiguous range \cite{radar}.
Hence, the effect of these $P$ targets on the transmitted AFDM signal can be represented by its discrete sampled version in the delay-time domain \cite{AFDM}
\begin{equation}
g_n[l]=\sum_{p=1}^{P} h_p e^{-j2\pi f_p n}\delta(l-l_p),\label{discretechan}
\end{equation}
where $f_p=(k_p+\kappa_p)/N_\mathrm{c}, k_p \in[-k_{p,\mathrm {max}},+k_{p,\mathrm {max}}]$ denotes the digital freequency normalized by sample points $N_\mathrm{c}$, $n$ and $l$ denote the time and delay indexes sampled at interval $1/B_\mathrm{c}$, respectively.
As Fig. \ref{AFDM-modulate} shows, considering an AFDM symbol containing $N_\mathrm{c}$ chirp subcarriers without phase coding (i.e., information symbols $x[m]$ are set to one), we can obtain the discrete time domain AFDM signal from the twisted time-frequency chirp domain \cite{ttfcdomain} by leveraging the inverse discrete affine Fourier transform (IDAFT)
\begin{equation}
{s}[n]=\frac{1}{\sqrt N_\mathrm{c}}\sum_{m=0}^{N_\mathrm{c}-1}e^{j 2 \pi\left(c_1 n^2+c_2 m^2+\frac{n m}{N_\mathrm{c}} \right)},\label{afdmmodulate}
\end{equation}
where $m,n=0,1,\cdots,N_\mathrm{c}-1$ denote the chirp subcarrier indices and time-domain sample indices respectively.


\subsection{Ambiguity Function and Parameters Selection}
To analyze the performance and characteristics of the waveform in greater detail, we introduce the concept of the ambiguity function. The ambiguity function serves as a metric to evaluate the sensing performance of a waveform, revealing the interference caused by a transmitted signal due to variations in any delay $\tau$ and Doppler shift $\nu$ compared to a reference signal. In this paper, the ambiguity function is defined as
\begin{equation}
\chi (\tau,\nu)=\int_{-\infty}^{+\infty}s^{\ast}(t)s(t+\tau)e^{j2\pi \nu t}dt.\label{abf}
\end{equation}

When we set $\nu=0$, the delay (time) dimensional versions of the ambiguity function can be obtained as
\begin{equation}
\chi (\tau)=\int_{-\infty}^{+\infty}s^{\ast}(t)s(t+\tau)dt,\label{delayabf}
\end{equation}
where $s(t)$ denotes the signal in the time domain, and $(\cdot)^*$ is the conjugate operation. To evaluate the discrete AFDM signal, the delay $\tau$ is sampled according to the interval $1/B_\mathrm{c}$
\begin{equation}
\chi [l]=\sum_{n=-\infty}^{+\infty}s^{\ast}[n]s[n+l],\label{discretedelayabf}
\end{equation}

First substituting \eqref{afdmmodulate} into \eqref{discretedelayabf}, we get

\begin{equation}
\begin{aligned}
\chi[l]=&\frac{1}{N_\mathrm{c}}\sum_{n=0}^{N_\mathrm{c}-1} \sum_{m=0}^{N_\mathrm{c}-1} \sum_{m^{\prime}=0}^{N_\mathrm{c}-1} e^{j 2\pi ( c_2 (m'^2 - m^2) }\\
&\times e^{j2\pi ( c_1 (n^2 - (n+l)^2) + \frac{m'n + m'l - mn}{N_\mathrm{c}})}.
\end{aligned}
\end{equation}

Then following the choice of $c_1 = \frac{2k_{{p,{p,\mathrm{max}}}}+1}{2N_\mathrm{c}}$ in \cite{AFDM}, which is the core parameter that enables AFDM to adapt to the time-varying channel without affecting the shape of the waveform in time domain, we have
\begin{equation}
\begin{aligned}
\chi[l]=&\frac{1}{N_\mathrm{c}}\sum_{n=0}^{N_\mathrm{c}-1} \sum_{m=0}^{N_\mathrm{c}-1}\sum_{m^{\prime}=0}^{N_\mathrm{c}-1}e^{j2\pi(- \frac{2k_{{p,{p,\mathrm{max}}}}+1}{N_\mathrm{c}}nl+\frac{m'l}{N_\mathrm{c}})}\\
&\times e^{j 2\pi \left( c_2 (m'^2 - m^2) + \frac{(m'-m)n}{N_\mathrm{c}} \right)}e^{j 2\pi(\frac{2k_{{p,\mathrm{max}}}+1}{2N_\mathrm{c}} l^2)}.
\end{aligned}
\end{equation}

When $m=m'$, it's clear that $\chi[l]$ contains most of the energy of the mainlobe due to the orthogonality between different AFDM subcarriers
\begin{equation}
\chi[l]=\frac{1}{N_\mathrm{c}}e^{j 2\pi(\frac{2k_{{p,\mathrm{max}}}+1}{2N_\mathrm{c}} l^2)}\sum_{n=0}^{N_\mathrm{c}-1} \sum_{m=0}^{N_\mathrm{c}-1}e^{j2\pi(- \frac{2k_{\mathrm {max}}+1}{N_\mathrm{c}}nl+\frac{m'l}{N_\mathrm{c}})},\label{m=m'}
\end{equation}
where we find that due to the summation range of $n$, $\chi[l] \equiv0$ unless $\chi[l]=1,l=0$.
But when $m\neq m'$, affected by the $e^{j 2\pi \left( c_2 (m'^2 - m^2) + \frac{(m'-m)n}{N_\mathrm{c}} \right)}$ term, $\chi[l]$ cannot ensure that it takes all zeros when $l\neq0$, and since different value of $c_2$ has distinct impacts on the phase relationship, each complex index changes and hence superimposes to produce diverse results.

From the point of view of optimizing $\chi[l]$, we will discuss three cases of $c_2$, i.e., the value near $\frac{1}{N_\mathrm{c}^2}$, the value far less than $\frac{1}{N_\mathrm{c}^2}$, and the value far larger than $\frac{1}{N_\mathrm{c}^2}$, respectively.
The reason why $\frac{1}{N_\mathrm{c}^2}$ is chosen as an intermediate value is that it has the same power as the squared difference term $m'^2-m^2$, which allows the periodicity of the exponent to be represented. The other two values are used in order to study the effect of the nonlinear term on the ambiguity function without considering the periodicity, which is a limit case analysis.

When $c_2$ is taken to be a real number near $\frac{1}{N_\mathrm{c}^2}$, the superposition of complex exponents will take on non-zero values when $l$ deviates from $0$ because the complex exponent corresponding to $m'^2-m^2$ can only take on squared terms within the period of $[0,N_\mathrm{c}^2-1]$ even though the complex exponent corresponding to the other terms can take on the full period.
However, due to the effect of periodicity, there are still a large number of complex exponential phases that cancel each other out, so the amplitude of $\chi[l]$ is not too high overall.

When $c_2$ is much larger than $\frac{1}{N_\mathrm{c}^2}$, the complex exponent corresponding to $m'^2-m^2$ will play a dominant role. When each complex exponent within the summation is not affected by the period $N_\mathrm{c}$, $\chi[l]$'s due to the rapid change of phase manifests itself as a rapid oscillation of the assignment, and the term $e^{j 2\pi(\frac{2k_{{p,\mathrm{max}}}+1}{2N_\mathrm{c}} l^2)}$ out of the summation makes its envelope decay to both sides.

When $c_2$ is much smaller than $\frac{1}{N_\mathrm{c}^2}$, we can ignore the effect of the quadratic term $e^{j 2\pi \left( c_2 (m'^2 - m^2) + \frac{(m'-m)n}{N_\mathrm{c}} \right)}$, and since $m'-m$ traverses the entire period, the superposition of periodic complex exponentials results in $0$ when $l\neq0$.

What's more, the ambiguity function in the frequency dimension for different parameterized AFDM presents the opposite of the time-domain ambiguity function, so that the choice of $c_2=1/N_\mathrm{c}^2$ will be a compromise between range and velocity estimation by AFDM. However, it is also because the AFDM with $c_2=0$ does not have the ability to measure velocity, its ranging performance will not be affected by target mobility, which makes our range estimation algorithm simple.

Therefore, the near-optimal time domain ambiguity function will be obtained when $c_2=0$, which can greatly improve the range sensing resolution and performance of the AFDM radar based on matched filtering. 

\begin{figure}[htbp]	
	\centering		\includegraphics[width=\linewidth,scale=1.00]{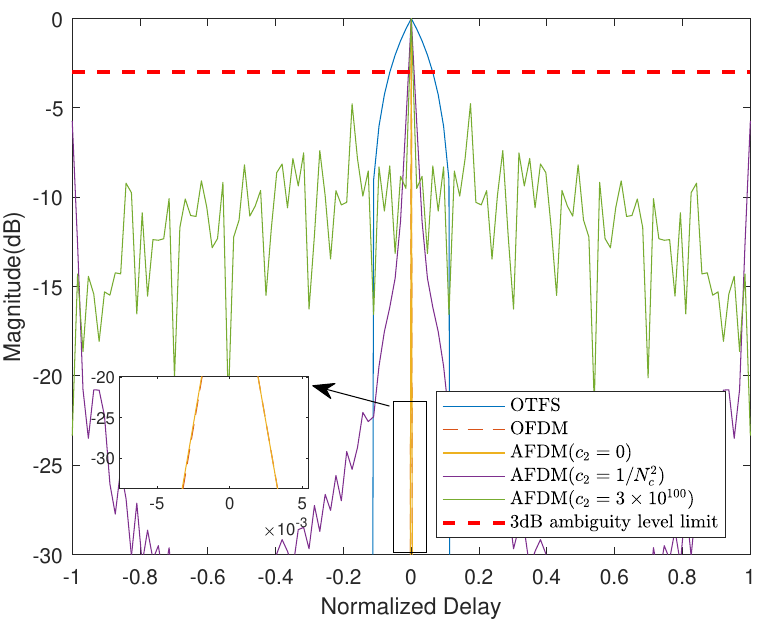}	
	\caption{Zero-Doppler cut of the ambiguity function.}	
	\label{AFDM-DELAYABF}	
\end{figure}
\begin{figure}[htbp]	
	\centering		\includegraphics[width=\linewidth,scale=1.00]{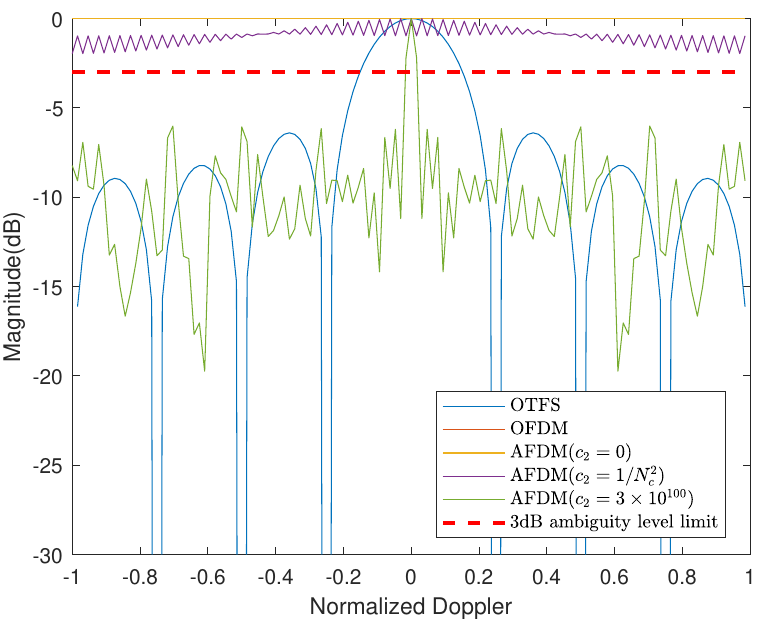}	
	\caption{Zero-delay cut of the ambiguity function.}	
	\label{AFDM-DOPLERABF}	
\end{figure}

\section{LOW-COMPLEXITY RANGE ESTIMATION ALGORITHM for CPP-AFDM}
\subsection{Range Estimation Algorithm}
In order to combat direct inter-copy interference between multiple received copies due to multipath effects, a CPP similar to the CP of OFDM is added before the AFDM signal $s[n],n=0,1,\cdots,N_\mathrm{c}-1$. But unlike the cyclic nature of OFDM, the AFDM signal exhibits a quasi-periodic nature, i.e., the cyclic shift leads to an additional phase $s[n+kN_\mathrm{c}]=e^{j2\pi c_1(k^2N_\mathrm{c}^2+2kN_\mathrm{c}n)}s[n]$.
Applying this quasi-periodic nature, we consider a $L_{\mathrm {CP}}$-length CPP whose length is greater than the number of samples of the channel's maximum delay expansion
\begin{equation}
s[n]=e^{-j2\pi c_1(N^2+2Nn)}s[n+N],\enspace n=-L_{\mathrm {CP}},\cdots,-1.\label{cppafdmmodulate}
\end{equation}

Finally, after adding the CPP to the original AFDM signal, the discrete-time domain AFDM received signal experiencing Doppler shift and multipath delay can be denoted as
\begin{equation}
r[n]=\sum_{l=0}^{+\infty} s[n-l] g_n[l]+w[n],\enspace n=-L_{\mathrm {CP}},\cdots,N_\mathrm{c}-1,\label{echo}
\end{equation}
where $w[n]\in \mathcal{CN}(0,N_\mathrm{0})$ is an additive white Gaussian noise (AWGN).
We note that when $2N_\mathrm{c}c_1$ is an integer and $N_\mathrm{c}$ is even, the CPP in \eqref{cppafdmmodulate} will degenerate to a simple CP (i.e., $s[n]=s[n+N],\enspace n=-L_{\mathrm {CP}},\cdots,-1$), so after removing the CPP in \eqref{echo}, the equivalent time-domain input-output relationship can be described as
\begin{equation}
r[n]=\sum_{l=0}^{+\infty} s[n-l]_{N_\mathrm{c}} g_n[l]+w[n],\enspace n=0,\cdots,N_\mathrm{c}-1,\label{equivalent in-output}
\end{equation}
where $[\cdot]_n$ denotes mod-$n$ operation. Due to the addition of the CPP, the interaction between the moving scatters (targets) and AFDM signal can be converted from linear in \eqref{echo} to circular convolution in \eqref{equivalent in-output}, which can be expressed as
\begin{equation}
r[n] = s[n]\star g[n]+w[n],\enspace n=0,\cdots,N_\mathrm{c}-1.\label{circonv}
\end{equation}
where $\star$ denotes circular convolution.
Traditionally, applying the classical time-domain pulse compression algorithm, with the input-output relationship of \eqref{echo}, we can obtain a target intensity vector
\begin{equation}
\hat g[n] = r[n]\ast s^{\ast}[n],\enspace n=-L_{\mathrm {CP}},\cdots,N_\mathrm{c}-1.\label{pulsecompression}
\end{equation}
where $\ast$ denotes linear convolution.
But for the time-domain transmit signal before adding CPP and the receive signal after removing CPP in \eqref{circonv}, \eqref{pulsecompression} can equivalently be written as
\begin{equation}
\hat g[n] = { r[n]} \star s^{\ast}[n],\enspace n=0,\cdots,N_\mathrm{c}-1.\label{circularpulsecompression}
\end{equation}

Then, according to the correspondence between the time-domain circular convolution and the frequency-domain product, we can use the low-complexity fast Fourier transform (FFT) and its inverse transform (IFFT) to put $\eqref{circularpulsecompression}$ in the frequency domain as
\begin{equation}
\hat g[n] = \mathrm{IFFT}( \mathrm{FFT}(r[n])\enspace \mathrm{FFT}( s^{\ast}[n])), n=0,\cdots,N_\mathrm{c}-1.\label{fftcircps}
\end{equation}

Finally, based on the constant false alarm rate equalizer (CFAR), the delay $\tau_p$ and its corresponding distance $R_p$ can be estimated for each target according to \eqref{range} and \eqref{rangesample}.

\subsection{Complexity Analysis}
In the previous section, we analyzed that the AFDM waveform possesses a near-optimal time-domain ambiguity function when $c_2=0$, and its characteristics are similar to those of OFDM, as shown in Fig. \ref{AFDM-DELAYABF} and \ref{AFDM-DOPLERABF}, enabling the simple pulse compression algorithm work without additional processing to combat target mobility.

According to the nature of matched filtering, we can realize it in the frequency domain with the help of low-complexity FFT. The complexity of implementing linear convolution directly in the time domain is $\mathcal O((N_\mathrm{c} + L_{\mathrm{CP}})^2))$ according to \eqref{pulsecompression}, the computational complexity after utilizing the FFT is $\mathcal O((N_\mathrm{c} + L_{\mathrm{CP}} ) \log(N_\mathrm{c} + L_{\mathrm{CP}}))$, since it is necessary to extend $r[n]$ and $s[n]$ to the same length. However, as for  \eqref{circularpulsecompression} and \eqref{fftcircps}, it will have lower complexity both directly in the time and frequency domains, the complexity of implementing the circular convolution in the time domain is $\mathcal O(N_\mathrm{c}^2)$, while the computational complexity after utilizing the FFT is $\mathcal O(N_\mathrm{c} \log(N_\mathrm{c})) $.



\begin{figure*}[t]
    \centering
    \subfigure[None of noise.]{\includegraphics[width=0.32\hsize, height=0.28\hsize]{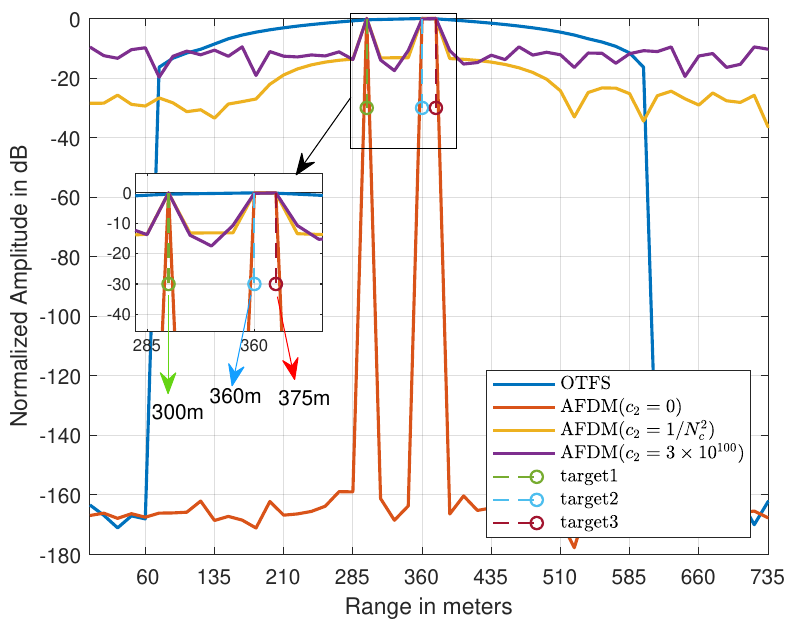}\label{AFDM_3_range_0noise}}
    \subfigure[Input SNR of 20 dB.]{\includegraphics[width=0.32\hsize, height=0.28\hsize]{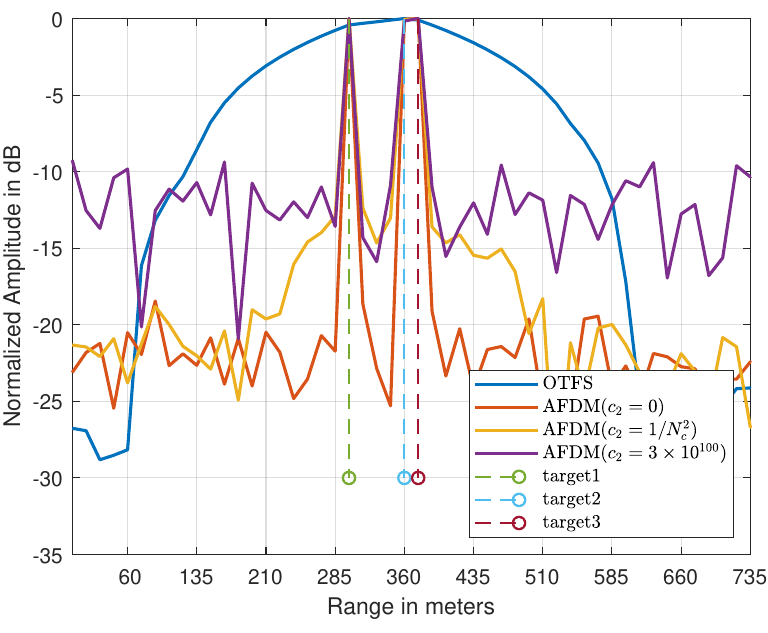}\label{AFDM_3_range_20dB}} 
    \subfigure[Input SNR of 0 dB.]{\includegraphics[width=0.32\hsize, height=0.28\hsize]{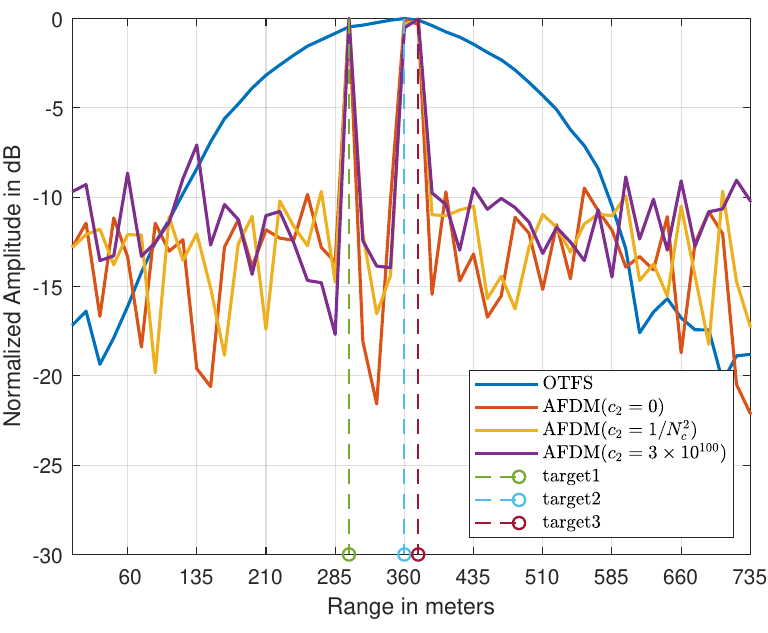}\label{AFDM_3_range_0dB}}
    \caption{AFDM with three $c_2$ parameters vs OTFS radar range profile in different noise environments with moving targets.}
    \label{range_estimation}
\end{figure*}

\begin{figure}[htbp]	
	\centering		\includegraphics[width=\linewidth,scale=1.00]{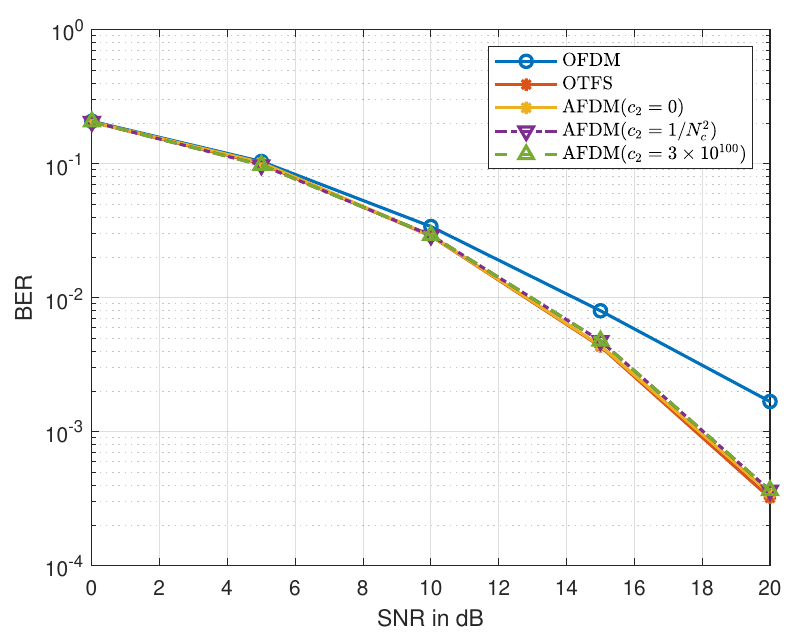}	
	\caption{BER performance of AFDM with three $c_2$ parameters vs OFDM and OTFS with LMMSE equalizer for QPSK.}	
	\label{AFDM-BER}	
\end{figure}

\section{SIMULATION RESULTS}
In this section, the system simulation parameters are: carrier frequency $f_\mathrm{c} = 24$GHz, subcarrier space $\Delta f = 39.063$kHz, number of AFDM subcarriers $N_{\mathrm{c}} = 256$, and number of AFDM symbols is $1$, so the bandwidth and time duration of the system are $N_{\mathrm{c}}\Delta f = 10$MHz and $1/\Delta f=25.6\mathrm{\mu s}$, respectively.
And comparison simulations are conducted in different noise environments to explore the range estimation performance of three AFDM waveforms with $c_2=0,1/256^2,3\times 10^{100}$ and OTFS with number of delay indices and Doppler indices equal to $\sqrt{N_\mathrm{c}}=16$, ensuring they have the same bandwidth and duration before adding CPP or reduced-CP \cite{PSOTFS}.
For the environment of scatters (targets), there are three point targets, at ranges $R_1 = 300$m, $R_2 = 360$m, and $R_3 = 375$m (with corresponding delays $\tau_1 = 2\mathrm{\mu s}, \tau_2 = 2.4\mathrm{\mu s}$ and $\tau_3 = 2.5\mathrm{\mu s}$), approaching the AFDM radar with velocities $v_1 = 24.4$m/s, $v_2 = 48.8$m/s and $v_3 = 122$m/s (corresponding Doppler shifts equal to $\nu_1 = 3.904$kHz, $\nu_2 = 7.808$kHz, and $\nu_3 = 19.52$kHz) respectively. Moreover, to ensure that the time duration of AFDM and OTFS waveforms is the same, the CPP and CP length $L_{\mathrm{CP}}$ are all the same. In addition, we apply the algorithm proposed in this paper at the radar sensing receiver to estimate the target range, and use AFDM demodulation, LMMSE equalization and constellation demapping at the communication receiver to estimate the communication information, respectively.

Fig. \ref{range_estimation} demonstrates the comparison of the range profiles of the three $c_2$ parameterized AFDM waveforms and the OTFS waveform in different noise environments. Fig. \ref{AFDM_3_range_0noise} is the ranging plots of each waveform for the ideal noiseless case, which corresponds to the zero-Doppler cut (i.e., time-dimentional) ambiguity function in Fig. \ref{AFDM-DELAYABF}. Since the energy of the AFDM with $c_2=0$ and OTFS is all concentrated in the main lobe area compared to the other waveforms, so that the energy of signals after matched filtered is also concentrated in the main lobe region, while the AFDM with the other two parameters accumulates considerable energy in the side lobes. However, OTFS has a wider main lobe compared to these three AFDM waveforms, which leads to lower range resolution, i.e., the peaks at several distances are too high to distinguish targets.
Therefore, since the AFDM with $c_2=0$ has the narrowest and most concentrated main lobe, its range estimation performance is the best among the four waveforms. For example, it can perfectly estimate the two targets immediately adjacent to each other at $360$ and $375$ meters, i.e., it achieves the ideal optimal range resolution capability. It is important to note that both targets can be detected using the threshold-based approach due to the consideration of integer delays, if fractional delays are considered (upsampling leads to more sampling points) there will be some lower values between these two targets and split it into two peaks. Moreover, compared to the other two parameterized AFDM waveforms, $c_2=0$ parameterized AFDM has very low energy except at the range of the target and thus does not cause false alarms due to high threshold settings under CFAR.

Fig. \ref{AFDM_3_range_20dB} and \ref{AFDM_3_range_0dB} show the range profiles of these four waveforms in environments with relatively low and high noise, respectively. Similar to the ideal noise-free environment, the four waveforms present the same range resolution capability, with the difference being that the noise leads to a rise in the magnitude of the sidelobe. It can be found that the amplitudes of the sidelobes of all waveforms in Fig. \ref{AFDM_3_range_0dB} increase more compared to those in Fig. \ref{AFDM_3_range_20dB} due to the larger noise, where the rise of the AFDM sidelobes for $c_2=0$ and $c_2=1/N_\mathrm{c}^2$ is obvious compared to that of the AFDM for $c_2=3\times10^{100}$, but it is still relatively lower. Therefore, it can be summarized that the AFDM with $c_2=0$ theoretically and practically has the best radar range performance and shows high robustness in an AWGN environment. Furthermore, we note that the ranging performance of AFDM under this parameter selection is not affected by target mobility, which can be explained by its frequency-dimensional ambiguity function as Fig. \ref{AFDM-DOPLERABF}.

In addition, we simulate the BER performance of AFDM under these three parameters and compare them to OTFS and OFDM under the time-varying multipath channel with Rayleigh fading. As Fig. \ref{AFDM-BER} shows, the three $c_2$ parameterized AFDM are almost the same as OTFS and have a lower BER than OFDM under the LMMSE equalizer, which makes the AFDM with $c_2=0$ a promising waveform for ISAC.

\section{Conclusion}
In this paper, we derive three AFDM waveforms' time-dimensional ambiguity functions with different built-in parameters $c_2$, and get the specific AFDM waveform with $c_2=0$, which has a near-optimal time-domain ambiguity function. 
Based on this understanding, we propose a low-complexity AFDM-based radar system that enables high-resolution ranging of moving targets, and both the waveform generation (modulation) and range estimation algorithms can be implemented with the help of FFT, enabling hardware multiplexing and real-time applications. Simulation results show that based on the low-complexity range estimation algorithm proposed in this paper, the specific AFDM not only possesses the best range discrimination capability and robustness against AWGN but also achieves the BER as low as OTFS.

\end{document}